\documentclass[%
 aip,
 amsmath,amssymb,
 reprint,%
]{revtex4-1}

\usepackage{graphicx}
\usepackage{dcolumn}
\usepackage{bm}

\usepackage[utf8]{inputenc}
\usepackage[T1]{fontenc}
\usepackage{mathptmx}
\usepackage{etoolbox}
\usepackage{epstopdf}
\usepackage{subfigure}
\usepackage[percent]{overpic}

\newcommand{\ket}[1]{|#1\rangle}
\newcommand{\bra}[1]{\langle #1|}

\newcommand{\n}{\nonumber\\}

\newcommand{\ex}[1]{\langle #1\rangle}

\makeatletter
\def\@email#1#2{%
 \endgroup
 \patchcmd{\titleblock@produce}
  {\frontmatter@RRAPformat}
  {\frontmatter@RRAPformat{\produce@RRAP{*#1\href{mailto:#2}{#2}}}\frontmatter@RRAPformat}
  {}{}
}%
\makeatother
\begin{document}

\preprint{AIP/123-QED}

\title[Approaching the double-Heisenberg scaling sensitivity in the Tavis-Cummings model ]{Approaching the double-Heisenberg scaling sensitivity in the Tavis-Cummings model }
\author{Yuguo Su}

\affiliation{ 
School of Science, Zhejiang University of Science and Technology, Hangzhou 310023, China
}%
\author{Tiantian Ying}%
\altaffiliation{yingtt@zjut.edu.cn}
\affiliation{ 
College of Civil Engineering, Zhejiang University of Technology, Hangzhou 310014, China
}%

\author{Bo Liu}
\altaffiliation{liuboheqingbo@126.com}
\affiliation{%
School of Information and Electrical Engineering, Hangzhou City University, Hangzhou 310015, China
}%

\author{Xiao-Guang Wang} 
\altaffiliation{xgwang@zstu.edu.cn}
\affiliation{Key Laboratory of Optical Field Manipulation of Zhejiang Province and Department of Physics, Zhejiang Sci-Tech University, Hangzhou 310018, China}


\date{\today}

\begin{abstract}
The pursuit of quantum-enhanced parameter estimations without the need for nonclassical initial states has long been driven by the goal of achieving experimentally accessible quantum metrology. 
In this work, employing a coherent averaging mechanism, we prove that the prototypical cavity quantum electrodynamics (QED) system, such as the Tavis-Cummings model, enables us to achieve not only the Heisenberg scaling (HS) precision in terms of the average photon number but also the double-HS sensitivity concerning both the average photon and atom numbers. 
Such a double sensibility can be experimentally realized by introducing either photon- or atom-number fluctuations through quantum squeezing.
Furthermore, we discuss the methodology to achieve this double-HS precision in a realistic experimental circumstance where the squeezing is not perfect.
Our results provide insights into understanding the coherent averaging mechanism for evaluating quantum-enhanced precision measurements and also present a usable metrological application of the cavity QED systems and superconducting circuits.
\end{abstract}

\maketitle

\section{Introduction}
Quantum metrology endeavors to achieve high-precision measurements of physical parameters by exploiting metrological resources, including quantum entanglement~\cite{PhysRevA.54.R4649,RevModPhys.90.035005,PhysRevLett.96.010401,PhysRevA.85.022321,PhysRevA.85.022322,PhysRevA.88.014301,PhysRevLett.126.080502,PhysRevLett.106.130506,Carollo_2019,10.21468/SciPostPhys.13.4.077,WOS:000970992400002} and quantum squeezing~\cite{PhysRevD.23.1693,RevModPhys.52.341,walls1983squeezed,VVDodonov_2002,PhysRevLett.97.011101,PhysRevLett.95.211102,PhysRevLett.88.231102,PhysRevLett.93.161105,PhysRevA.92.023603}, etc.
Remarkable progress in experimental technology has facilitated extensive applications of quantum metrology in a variety of fields such as atomic clocks~\cite{RevModPhys.87.637,Louchet_Chauvet_2010,PhysRevLett.112.190403}, gravitational wave detectors~\cite{Abbott_2009}, magnetometry~\cite{budker_optical_2007,PhysRevLett.109.253605,PhysRevLett.120.260503,PhysRevLett.126.010502,PhysRevA.109.042614,PhysRevX.5.031010} and quantum imaging~\cite{Genovese_2016,PhysRevA.95.063847,PhysRevA.96.062107,deng2023,tan2023quantum}.
A central challenge of quantum metrology involves identifying inherent metrological limitations and designing accessible schemes to enhance the precision of parameter estimation.
For this purpose, it is necessary to analyze the scaling behavior of the quantum Fisher information (QFI), which poses the theoretical limit for estimating an unknown parameter $\gamma$ via the quantum Cram\'er-Rao bound~\cite{PhysRevLett.72.3439} $\delta \gamma\geq1/\sqrt{\nu\mathcal{F}_{\gamma}}$.
This bound can be asymptotically approached by increasing the number of independent measurements $\nu$.

In the scenario of the quantum phase estimation, the highest achievable precision using the separable $N$-particle states is known as the standard quantum limit (SQL), represented by $\mathcal F_{\gamma}\propto N$.
However, the entangled states are indispensable for surpassing this limit to approach the Heisenberg limit (HL), i.e., $\mathcal F_{\gamma}\propto N^2$.
Instead of directly employing entangled probes, numerous alternative approaches have been explored to enhance measurement precisions by 
harnessing the quantum features arising from many-body physics~\cite{PhysRevLett.130.170801,PhysRevLett.132.100803}, such as criticality-enhanced metrology~\cite{PhysRevLett.126.010502,PhysRevB.109.L041301,PhysRevLett.99.095701,PRXQuantum.3.010354,PhysRevLett.126.200501,PhysRevLett.126.200501,PhysRevLett.121.020402,PhysRevA.78.042105,PhysRevLett.124.120504}, chaotic quantum metrology~\cite{fiderer2018quantum,PhysRevA.103.023309,li2023improving}, quantum Zeno effect-based metrology~\cite{PhysRevLett.129.070502}, or by optimal adaptive quantum controls~\cite{pang_optimal_2017,PhysRevA.96.020301,PhysRevLett.128.160505}.

A promising scheme called ``coherent averaging''~\cite{braun_heisenberg-limited_2011,https://doi.org/10.1002/andp.201500169,RevModPhys.90.035006} has been proposed, where a central spin (a ``quantum bus'') is coupled with a cavity.
Thus, it becomes possible to utilize an initial product state to attain the Heisenberg scaling (HS) sensitivity for estimating a global phase.
Specifically, the core of the coherent averaging mechanism lies in the global parameter $\gamma$, which serves as the prefactor of the density-density coupling Hamiltonian, i.e., the encoding Hamiltonian in the form of $\exp(-i\gamma H)$, where $H$ represents the density-density coupling.
Consequently, a crucial question naturally arises: Can we improve measurement precision by enlarging the quantum bus from a single atom to $N$ atoms?
Moreover, is it feasible to extend the coherent averaging mechanism for precise estimation of a non-global parameter?

\begin{figure}[t]
	\centering
	\begin{minipage}{1\linewidth}
		\begin{overpic}[width=\linewidth]{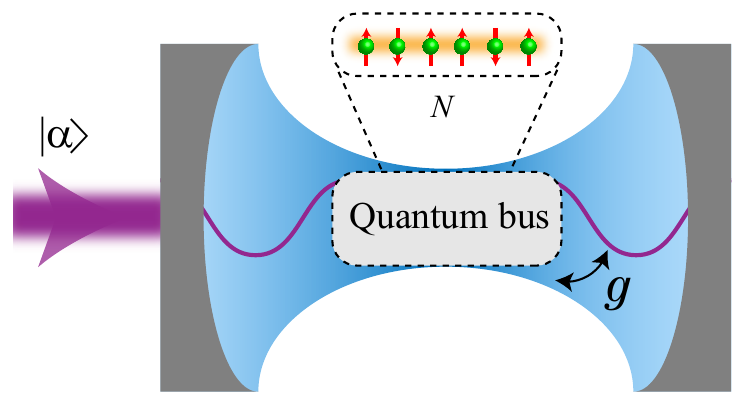}
		\end{overpic}
	\end{minipage}
	\caption{Cavity-QED setup.
		A coherent field $\left|\alpha\right\rangle$ with $\bar n$ photons is injected into the cavity and couples to $N$ trapped two-level atoms (green) with the coupling strength $g$.
		The $N$ atoms, serving as a quantum bus, exhibit equal coupling strengths to the optical field, forming the TC model.
		We will estimate a non-global parameter, the strength of a weak magnetic field $h$, appearing in the Hamiltonian~\eqref{H-1}.
	}
	\label{Fig1-TC-Model}
\end{figure}

In this work, we report on a realization of the double-HS sensitivity $\mathcal F_{\gamma}\propto N^2 \bar{n}^2$ with respect to the number of two-level atoms $N$ and the average number of photons $\bar{n} $ in the Tavis-Cummings (TC) model~\cite{PhysRev.170.379}, see Fig.~\ref{Fig1-TC-Model}. 
This model provides a versatile platform for investigating the interplay between light and matter in the paradigm of quantum optics and quantum metrology. 
In contrast to the previous scenario with a single central spin, the $N$ atoms in the TC model, acting as a quantum bus, could show the collective interplay with the light and involve quantum resources.
We claim that the coherent averaging mechanism naturally emerges, enabling the estimation of even a non-global parameter, such as a weak magnetic field, in the TC model. 
Furthermore, we demonstrate that the introduction of quantum fluctuations in photon or atom numbers, achieved through squeezing, allows for the attainment of double-HS sensitivity.
Our findings highlight a potential application of the TC model to the experimental implementation of the high-precision quantum measurement with experimentally accessible quantum resources.

In Sec. \ref{SEC2}, we present the TC model, and employ the time-averaged method to derive an effective description of the TC model.
In Sec. \ref{SEC3}, we introduce the metrological scheme and derive the analytic QFI to quantify measurement precision.
The impact of selecting different initial states on QFI is comprehensively discussed in Sec. \ref{SEC4}.
Finally, Sec. \ref{SEC5} summarizes and discusses our findings.

\section{Tavis-Cummings Model and its effective description}\label{SEC2}
The TC model characterizes the coupling of a single bosonic cavity mode to a collection of $N$ two-level atoms (or spin-1/2 spins), as illustrated in Fig.~\ref{Fig1-TC-Model}:
\begin{align}
	H&=(h+\omega_0)J_z+\omega_{\text{a}} a^\dagger a+H_{\rm I},\label{H-1}\\
	H_{\rm I}&=g\left(a^\dagger J_-+aJ_+\right),\nonumber
\end{align}
where $H_{\rm I}$ denotes the interaction Hamiltonian and $J_{x,y,z}=\sum_{j=1}^{N}{\sigma_j^{x,y,z}/2}$, $J_\pm=J_x\pm \text{i}J_y$ are the collective spin operators; $a^\dagger$ and $a$ are creation and annihilation operators for the cavity mode;  $\omega_0$, $\omega_{\text{a}}$ and $2g$ are the original spin transition, the cavity frequency and the single-photon Rabi frequency; $h$ is the magnetic field to be estimated.
Throughout the paper, we set the Planck constant $\hbar=1$.
Next, we will derive an effective metrological Hamiltonian for estimating the weak magnetic field $h$ and prove that the coherent averaging mechanism will emerge.

After applying the unitary transformation $U_{1}=\exp\left[{-\mathrm{i}\left(\omega J_z+\omega_{\text{a}} a^\dagger a\right)t}\right]$, the total Hamiltonian~\eqref{H-1} in the interaction picture takes the form:
\begin{equation}
	H^{(\text{I})} = \text{i}\frac{dU_{1}^{\dagger}}{dt}U_{1}+U_{1}^{\dagger}HU_{1}=g\left(a^{\dagger}J_{-}\text{e}^{-\text{i}\Delta t}+\text{H.c.}\right),
\end{equation}
where $\Delta=\omega-\omega_{a}$ is the effective detuning and $\omega\equiv \omega_0+h$.
Utilizing the time-averaged method of the Ref.~\cite{James2007}, the interaction Hamiltonian $H^{(\text{I})}=\sum_{i=1}^{N}\left(f_{i}\text{e}^{-\text{i}\Delta_{i}t}+\text{H.c.}\right)$ could be rewritten as the following compact form
\begin{align}
	H^{(\text{I})} &\approx \sum_{i,j=1}^{N}\frac{1}{\bar{\Delta}_{ij}}\left[f_{i}^{\dagger},f_{j}\right]\text{e}^{\text{i}\left(\Delta_{i}-\Delta_{j}\right)t}\nonumber\\
	&=-\frac{g^{2}}{\Delta}\left[a^{\dagger}J_{-},aJ_{+}\right]\nonumber\\
	&=\frac{g^{2}}{\Delta}\left(2J_{z}a^{\dagger}a+J_{+}J_{-}\right),\label{TAM}
\end{align}
where the operator $f_{i}=ga\sigma_{i}^{+}/2$, the $i$-th effective detuning $\Delta_i\equiv\Delta$ and the harmonic average frequency $\bar{\Delta}_{ij}\equiv\left|2\Delta_{i}\Delta_{j}/\left(\Delta_{i}+\Delta_{j}\right)\right|=\left|\Delta\right|>0$ correspond to our system.
Alternatively, the Fr{\"o}lich-Nakajima transform is performed to obtain our effective Hamiltonian~\cite{MA201189}, and the validity of the approximation in Eq.~\eqref{TAM} is guaranteed by the large-detuning condition, i.e., $\Delta\gg g\sqrt{N}$.
Returning to the Schr\"{o}dinger picture, we obtain the effective Hamiltonian as follows
\begin{align}\label{H^s_eff}
	H_{\text{eff}}^{\left(\text{s}\right)}&=\text{i}\frac{dU_{1}}{dt}U^{\dagger}_{1}+U_{1}H^{(\text{I})}U^{\dagger}_{1} \nonumber\\
	&\simeq \omega J_{z}+\omega_{a}a^{\dagger}a+\frac{2g^{2}}{\Delta}J_{z}a^{\dagger}a+\frac{g^{2}}{\Delta}J_{+}J_{-},
\end{align}
Generally, the average photon number $\bar{n}\equiv\left\langle a^{\dagger}a\right\rangle $ is much larger than the number of up spins, i.e., $\bar n\gg\ex{J_z}= \ex{[J_+,J_-]}/2$, then the Hamiltonian~\eqref{H^s_eff} can be further reduced as follows: 
\begin{equation}
	H_{\text{eff}}=\omega J_z+\omega_{\rm{a}} a^\dagger a+\frac{2g^2}{\Delta}J_za^\dagger a.\label{H_eff}
\end{equation}

It should be emphasized that the estimated parameter $h$ now appears in the coupling term, i.e., $\Delta=\omega_0-\omega_{\text{a}}+h$, which reveals the existence of the coherent averaging mechanism in the TC model.
The cavity QED system with strong light-atom interactions and large detuning will suppress the atomic spontaneous emission $\gamma_e$ and the cavity dissipation $\gamma_d$.
Furthermore, due to continuous experimental efforts to realize strong coupled cavity QED, we only consider the effective Hamiltonian~\eqref{H_eff} without dissipation case~\cite{WOS:000243867300038,WOS:000322086100035,PhysRevA.71.013817,PhysRevA.67.033806,WOS:000257665300034,PhysRevLett.128.123602,PhysRevLett.130.173601,PhysRevLett.103.083601}. 
We will demonstrate that the Hamiltonian~\eqref{H_eff} enables achieving  $\bar n^2$ HS sensitivity without requiring initial entanglement or squeezing between atoms and photons due to the emergent coherent averaging mechanism, i.e., the effective coupling term $J_za^\dag a$.
We will also show an attainable double-Heisenberg-scaling sensibility in measuring the weak field $h$ by introducing some quantum resources.

\section{Metrological scheme and Quantum Fisher information}\label{SEC3}
We consider an initial probe state of a bipartite product of an optical component and an atomic component, i.e., $\rho=\rho_{\rm op}\otimes \rho_{\rm at}$.
The optical component is chosen as a single-mode Gaussian state~\cite{WANG20071,RevModPhys.84.621}
\begin{align}
	\rho_{\rm op}=D\left(\alpha\right)S\left(\xi\right)\rho_{\text{th}}S^{\dagger}\left(\xi\right)D^{\dagger}\left(\alpha\right),\label{DSTS}
\end{align}
which gives a displaced squeezed thermal state (DSTS). 
In the above equation, $\rho_{\text{th}}=\sum_{n}p_{n}\left|n\right\rangle \left\langle n\right|$ denotes a thermal state with a probability amplitude $p_{n}=n_{\text{th}}^{n}/\left(1+n_{\text{th}}\right)^{n+1}$, where $n_{\text{th}}$ is the thermal photon number and $\left|n\right\rangle $ denotes the Fock state. 
While $D\left(\alpha\right)=\exp\left(\alpha a^{\dagger}-\alpha^{*}a\right)$ is the displacement operator with a displacement parameter $\alpha=\left|\alpha\right|\exp\left(\text{i}\zeta\right)$ ($\left|\alpha\right|\geq0$) and $S\left(\xi\right)=\exp\left(-\xi a^{\dagger2}/2+\xi^{*}a^{2}/2\right)$ is a squeezing operator with a squeezing parameter $\xi=r\exp\left(\text{i}\vartheta\right)$ ($r\geq0$).
In general, the optical quantum state $\rho_{\rm op}$ can characterized by the parameters $\left(\left|\alpha\right|,\zeta,r,\vartheta,n_{\text{th}}\right)$, which represent the displacement amplitude, the displacement angle, the squeezing amplitude, the squeezing angle, and the thermalization, respectively. 

Following the unitary evolution $\rho(t)=U(t)\rho U^\dag(t)$ where $U(t)=\text{e}^{-\mathrm{i}H_{\text{eff}}t}$, the estimated parameter $h$ will be encoded into the evolved state $\rho(t)$.
The precision associated with estimating the magnetic field $h$ is governed by the quantum Cram\'er-Rao bound~\cite{PhysRevLett.72.3439} 
\begin{equation}\label{C-R bound}
	\delta h\geq\frac{1}{\sqrt{\mathcal F_h}},
\end{equation}
where the QFI can be determined by the following formula~\cite{Liu_2019, YuguoSu2021}
\begin{equation}
	\mathcal{F}_h\!=\!\sum_{p_i\in\mathcal{S}}{\!4p_i\!\left\langle\Psi_i\!\left|\mathcal{H}_h^2\right|\!\Psi_i\!\right\rangle}-\!\sum_{p_i\!,p_j\!\in\mathcal{S}}{\frac{8p_ip_j}{\!p_i\!+\!p_j\!}\!\left|\left\langle\Psi_i\!\left|\mathcal{H}_h\right|\!\Psi_j\!\right\rangle\right|^2\!}.\label{QFI}
\end{equation}
Here, $\left|\Psi_{i}\right\rangle $ is $i$-th eigenstate of the initial system $\rho=\rho_{\rm op}\otimes \rho_{\rm at}$ with the eigenvalue $p_{i}$, $\mathcal{S}=\left\{ p_{i}\in\left\{ p_{i}\right\}|p_{i}\neq0\right\}$ is the support of the initial density matrix.
Based on the effective Hamiltonian~\eqref{H_eff}, the generator of quantum sensing is given by 
\begin{equation}\label{Generator-QS}
	\mathcal{H}_h  = \mathrm{i}[\partial_hU^{\dagger}(t)]U(t)=\left(\frac{2g^2}{\Delta^2}a^\dagger a-1\right)tJ_z.
\end{equation}
We restrict the initial optical cavity to a Gaussian state~\eqref{DSTS} due to its encompassment of commonly encountered quantum states in experiments, such as the coherent and squeezed states.
By substituting the initial system $\rho=\rho_{\rm op}\otimes\rho_{\rm at} $~\eqref{DSTS} and the generator~\eqref{Generator-QS} of quantum sensing into the definition of the QFI~\eqref{QFI}, we obtain the QFI as follows
\begin{align}
	\mathcal{F}_h&=4t^{2}\left\{ \left(1-\frac{2g^{2}}{\Delta^{2}}\bar{n}\right)^{2}\text{Var}\left(J_{z}\right)+\frac{4g^{4}}{\Delta^{4}}\text{Var}\left(a^{\dagger}a\right)\left\langle J_{z}^{2}\right\rangle\right.\nonumber\\
	&\quad\left.-\frac{4g^{4}}{\Delta^{4}}n_{\text{th}}\left(n_{\text{th}}+1\right)\left\langle J_{z}\right\rangle ^{2}\left[\frac{\cosh4r\left(2n_{\text{th}}+1\right)^{2}+1}{4n_{\text{th}}\left(n_{\text{th}}+1\right)+2}\right.\right.\nonumber \\
	&\quad\left.\left. +\frac{4\left|\alpha\right|^{2}}{2n_{\text{th}}+1}\left[1+2\sinh^{2}r-\sinh2r\cos\left(2\zeta-\vartheta\right)\right]\right]\right\}, \label{F_DSTS}
\end{align}
where
\begin{align}
	\text{Var}\left(a^{\dagger}a\right)
	&=\frac{\sinh^{2}2r}{2}\left(2n_{\text{th}}^{2}+2n_{\text{th}}+1\right)\nonumber\\
	&\quad+ \left|\alpha\right|^{2}\!\left(2n_{\text{th}}\!+\!1\right)\left[\cosh\!2r\!-\!\sinh\!2r\cos\left(2\zeta\!-\!\vartheta\right)\right]\nonumber\\
	&\quad +\left(1+2\sinh^{2}r\right)^{2}n_{\text{th}}\left(n_{\text{th}}+1\right)\label{Vara+a},
\end{align}
is the photon-number fluctuation, $\left\langle\cdot\right\rangle$ represents the expectation with respect to the initial state, and $\text{Var}\left(O\right)=\left\langle O^{2}\right\rangle -\left\langle O\right\rangle ^{2}$ is the variance of the operator $O$. 
Subsequently, we will proceed to demonstrate that the analytical outcome \eqref{F_DSTS} for the QFI offers highly practical theoretical insights for metrological applications within the cavity QED systems.

\section{Results}\label{SEC4}
\subsection{Coherent spin state with different photon states}
In this section, we fix the atom state to be uncorrelated, namely separable states, and explore the conditions that allow a photon state to achieve a double-HS precision.
Separable spin-coherent states $\rho_{\rm at}=\ket{\mu}_{\rm at\ at}\bra{\mu}$ are given by~\cite{JMRadcliffe_1971}
\begin{align}
	\left|\mu\right\rangle \! & =\! \frac{\exp\!\left(\!\mu\! J_{-}\!\right)}{\left(\!1\!+\!\left|\mu\right|^{2}\!\right)^{j}}\left|0\right\rangle \!=\!\frac{1}{\left(\!1\!+\!\left|\mu\right|^{2}\!\right)^{j}}\sum_{p=0}^{2j}\sqrt{\!\frac{\left(2j\right)!}{p!\!\left(\!2j\!-\!p\!\right)!}\!}\mu^{p}\!\left|p\right\rangle,\label{SCS}
\end{align}
where $\left| p \right\rangle \equiv \left| j,j-p \right\rangle $ with $\left|j,j\right\rangle =\left|0\right\rangle $, being the eigenstate of $J_{z}$ for $j=N/2$.
Meanwhile, the phase parameter $\mu$ is parameterized by the angles $\left(\theta,\phi\right)$ via the stereographic projection $\mu=\text{e}^{\text{i}\phi}\tan\left(\theta/2\right)$ [$\theta\in\left[0,\pi\right)$ and $\phi\in\left[0,2\pi\right)$].
In order to ascertain the separability of $\ket{\mu}$, we can rewrite it as 
\begin{equation}
	\ket{\mu}\equiv\ket{\theta,\phi}=\bigotimes_{i=1}^N\left(\cos\frac{\theta}{2}\ket{\uparrow}_i+e^{{\rm i}\phi}\sin\frac{\theta}{2}\ket{\downarrow}_i\right),
\end{equation}
which is expressed as the tensor product of identical qubits, thereby leading us to categorize it as a classical atom state. 

\emph{Optical coherent state}.---Firstly, we consider the photon state as a classical coherent state (CS) $\ket{\psi}_{\rm op}=\ket{\alpha}=D(\alpha)\ket{0}$, where $\ket{\psi}_{\rm op}$ is the wavefunction of the reduced density matrix $\rho_{\rm op}=\ket{\psi}_{\rm op\ op}\bra{\psi}$.
From Eqs.~\eqref{F_DSTS} and \eqref{SCS}, we obtain 
\begin{align}
	\mathcal{F}_h^{\text{CS}} 
	&= 4t^{2}\!\left[\left(\!1\!-\!\frac{2g^{2}}{\Delta^{2}}\bar{n}\!\right)^{2}\!\text{Var}\!\left(J_{z}\right)\!+\!\frac{4g^{4}}{\Delta^{4}}\text{Var}\!\left(a^{\dagger}a\right)\left\langle J_{z}^{2}\right\rangle \right]\nonumber \\
	& =   4t^{2}\!\left[\left(\!1\!-\!\frac{2g^{2}}{\Delta^{2}}\bar{n}\!\right)^{2}\!\frac{N}{4}\!\sin^{2}\!\theta\!+\!\frac{g^{4}}{\Delta^{4}}\bar{n}N\!\left(\sin^{2}\!\theta\!+\!N\!\cos^{2}\!\theta\right)\right],\label{QFI_CS2}
\end{align}
where $n_{\text{th}}=0$,  $\bar{n}=\left|\alpha\right|^{2}$, $\text{Var}\left(J_{z}\right)=\left(N\sin^{2}\theta\right)/4$ and $\text{Var}\left(a^{\dagger}a\right)=\bar{n}$.
Under the large average photon number condition ($\bar{n}\gg N$ and $2g^{2}\bar{n}/\Delta^{2}\gg1$),
the above expression of QFI becomes 
\begin{equation}
	\mathcal{F}_h^{\text{CS}} 
	\approx  t^{2}\frac{16g^{4}}{\Delta^{4}}\bar{n}^{2}\text{Var}\left(J_{z}\right) 
	=\frac{4g^{4}}{\Delta^{4}}t^{2}N\bar{n}^{2}\sin^{2}\theta, \label{QFI_CS3}
\end{equation}
that suggests the maximum QFI $\mathcal{F}_h^{\text{CS}}\approx4g^{4}t^{2}N\bar{n}^{2}/\Delta^{4}$ for $\theta=\pi/2$. 
Hence, the result~\eqref{QFI_CS3} gives the HS precision ($\propto \bar n^2$) through the utilization of the coherent averaging mechanism without using any entanglement of the initial state.
In particular, the initial probe states are totally classical in the sense that the photon state is a coherent state and the atom state is a separable spin-coherent state. 

The QFI provides the optimal measurement precision by optimizing over all positive-operator-valued measure measurements.
However, not all the experimental measurements can reach the precision limit bounded by the QFI, i.e., $\delta h \simeq 1/\sqrt{\mathcal F_{h}}$. 
The measurement precision $\delta h$ is given by the error-propagation formula
\begin{equation}
	\delta^2h=\frac{{\rm Var}[M(t)]}{\left|\partial_h\left\langle M(t)\right\rangle\right|^2},\label{Error-formula}
\end{equation}
where $M$ is an observable to be determined.
In our case, we choose the $M=J_{\varphi}\equiv J_{x}\cos\varphi+J_{y}\sin\varphi$.
By using the Baker-Campbell-Hausdorff formula~\cite{Baker1901} and the effective Hamiltonian~\eqref{H_eff}, we find 
\begin{align}
	M(t)&=\text{e}^{\text{i}tH_{\text{eff}}}M\text{e}^{-\text{i}tH_{\text{eff}}}=  \frac{1}{2}\left[J_{+}\text{e}^{-\text{i}(\varphi-\hat{\nu} t)}+J_{-}\text{e}^{\text{i}(\varphi-\hat{\nu} t)}\right],\n 
	M^2(t)
	&= \frac{1}{4}\left[\!N\!\left(\!\frac{N}{2}\!+\!1\!\right)\!-\!2J_{z}^{2}\!+\!J_{+}^{2}\text{e}^{-2\text{i}(\varphi-\hat{\nu} t)}\!+\!J_{-}^{2}\text{e}^{2\text{i}(\varphi-\hat{\nu} t)}\right],\nonumber
\end{align}

where 
$\hat{\nu}\equiv\omega_{0}+h+2g^{2}a^{\dagger}a/\Delta$.
For an initial spin-coherent state $\ket{\theta,\phi}$, Eq.~\eqref{SCS},  the expectation values are given by
\begin{align}
	\left\langle M(t)\right\rangle
	&= \frac{N}{2}\sin\theta\cos(\nu t+\phi-\varphi),\label{M1}	\\
	\left\langle M^{2}(t)\right\rangle
	& =  \frac{1}{16}N\left\{ N+3+\left(1-N\right)\cos2\theta\right.\n
	&\quad\left.+2(N-1)\sin^{2}\theta\cos\left[2\left(\nu t+\phi-\varphi\right)\right]\right\},\label{M2}	
\end{align}
where 
$\nu\equiv\omega_{0}+h+2g^{2}\bar n/\Delta$.
By substituting Eqs.~\eqref{M1} and~\eqref{M2} into the error-propagation formula~\eqref{Error-formula}, we have 
\begin{align}
	\left(\delta h\right)^{2} 
	& =  \frac{\csc^{2}\theta\csc^{2}\left(\nu t\!+\!\phi\!-\!\varphi\right)\!-\!\cot^{2}\left(\nu t\!+\!\phi\!-\!\varphi\right)}{Nt^{2}\left(\frac{2g^{2}}{\Delta^{2}}\bar{n}-1\right)^{2}}\label{FI_1}\\
	&\gtrsim \frac{\csc^{2}\theta\csc^{2}\left(\nu t\!+\!\phi\!-\!\varphi\right)\!-\!\cot^{2}\left(\nu t\!+\!\phi\!-\!\varphi\right)}{4t^{2}\frac{g^{4}}{\Delta^{4}}N\bar{n}^{2}},
\end{align}
when the average photon number satisfies $2g^{2}\bar{n}/\Delta^{2}\gg1$.
By choosing $\theta=\pi/2$, the greatest lower bound is taken as follows
\begin{equation}
	\left(\delta h\right)^{2}  \gtrsim \frac{1}{4t^{2}\frac{g^{4}}{\Delta^{4}}N\bar{n}^{2}}.
\end{equation}

\begin{figure}
	\centering
	\begin{minipage}{1\linewidth}
		\begin{overpic}[width=\linewidth]{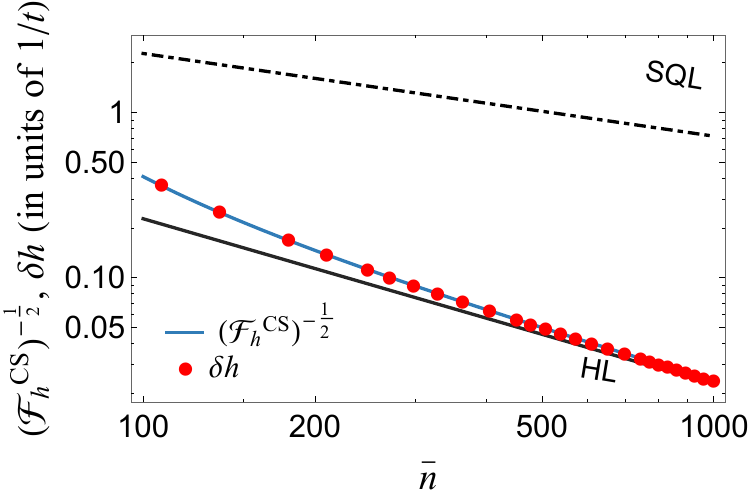}
		\end{overpic}
	\end{minipage}
	\caption{QFI $\mathcal{F}_{h}^{\text{CS}}$ and measurement precision $\delta h$ (in units of 1/$t$) of the cavity QED system versus the average photon number $\bar{n}$, with the spin number $N=4$, the original atomic transition frequency $\omega_0/\left(2\pi\right)=6.9$ GHz, the cavity frequency $\omega_{\text{a}}/\left(2\pi\right)=6.89$ GHz, the Rabi frequency $g/\left(2\pi\right)=1.05$ MHz, the frequency equivalence of the magnetic field $h=0.1$ mHz, the rotation angle $\theta=\pi/2$.
		The blue line stands for the QFI $\mathcal{F}_{h}^{\text{CS}}$ obtained from Eq.~\eqref{QFI_CS2}; meanwhile, the red circles denote the measurement precision $\delta h$ obtained from numerical calculation via the error-propagation-formula.
		The black dotted-dashed line is for SQL, and the black line denotes the HL.
		Our setting refers to the experimental parameters given in Ref.~\cite{WOS:000243867300038}.
	}\label{Fig2-HL-CS}
\end{figure}

Figure~\ref{Fig2-HL-CS} represents the QFI $\mathcal{F}_{\omega}^{\text{CS}}$ and measurement precision $\delta \omega$ (units of 1/$t$) of the cavity QED system versus the average photon number $\bar{n}$.
It shows that the optimal measurement precision determined by the QFI~\eqref{QFI_CS2}, i.e., the HS precision, can be achieved by performing measurements on the spin part.
One could find that measurement precision $\delta \omega$ (red dashed line) gained from Eq.~\eqref{FI_1} is bounded by the QFI $\mathcal{F}_{\omega}^{\text{CS}}$ obtained from Eq.~\eqref{QFI_CS2} (cyan line), which is corresponded with the QCRB.
Moreover, both are beyond the SQL (black dot dash line) and approach the HL (black line) with a large average photon number $2g^{2}\bar{n}/\Delta^{2}\gg1$.


\emph{Optical squeezed vacuum state}.---In the second case, we replace the classical coherent photon state with a squeezed vacuum state (SVS) $\ket{\psi}_{\rm op}=\left|\xi\right\rangle\equiv S\left(\xi\right)\left|0\right\rangle $.
By invoking Eq.~\eqref{F_DSTS} with $|\alpha |^{2}=n_{\text{th}}=0$, $\text{Var}\left(J_{z}\right)=0$ and $\left\langle \mu\right|J_{z}^{2}\left|\mu\right\rangle =N^{2}/4$ for $\theta=0$, the associated QFI is given by 
\begin{equation}\label{O-SVS}
	\mathcal{F}_h^{\text{SVS}} 
	\!=\! t^{2}\frac{16g^{4}}{\Delta^{4}}\text{Var}\left(a^{\dagger}a\right)\left\langle J_{z}^{2}\right\rangle 
	\!=\! t^{2}\frac{2g^{4}}{\Delta^{4}}N^{2}\sinh^{2}2r.
\end{equation}
For perfect squeezing, $\bar{n}=\sinh^{2}r$,  and considering substantial quantum fluctuation such that $\text{Var}\left(a^{\dagger}a\right)=2\sinh^{2}r\cosh^{2}r\geq2\bar{n}^{2}$, we further approximate Eq.~\eqref{O-SVS} as 
\begin{equation}
	\mathcal{F}_h^{\text{SVS}} \geq  \frac{8g^{4}}{\Delta^{4}}t^{2}N^{2}\bar{n}^{2},
\end{equation}
which implies the attainment of the double-HS precision ($\propto N^2\bar n^2$) through the utilization of an SVS as the only resource state. 
We observe that the double-HS precision elucidated by Eq.~\eqref{O-SVS} is essentially attributed to the fluctuation in photon number, exhibiting the scaling of $\bar n^2$ induced by the SVS. 
Consequently, the double-HS contribution exists in the QFI $\mathcal{F}_h^{\text{SVS}} \propto \text{Var}\left(a^{\dagger}a\right)\left\langle J_{z}^{2}\right\rangle \propto \bar n^2N^2$, regardless of whether the atom state is classical or nonclassical.
This presents an avenue for achieving the double-HS precision, accomplished by harnessing the photon-number fluctuations engendered by squeezed photon states.

\begin{figure}[t]
	\centering
	\begin{minipage}{1\linewidth}
		\begin{overpic}[width=\linewidth]{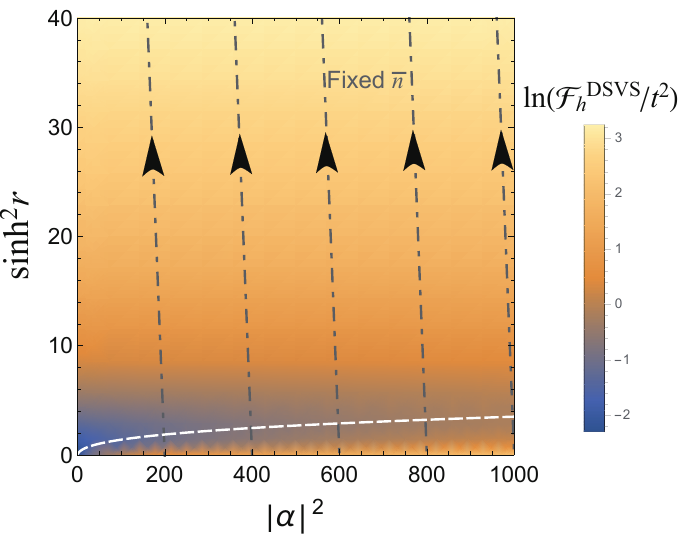}
		\end{overpic}
	\end{minipage}
	\caption{Density plot of the rescaled QFI $\ln\left(\mathcal{F}_{h}^{\text{DSVS}}/t^{2}\right)$ as a function of displacement $\left|\alpha\right|^{2}$ and squeezing $\sinh^{2}r$ with the angle condition $\tau=1$.
		The grey dot-dashed lines denote the states with a given average photon number (i.e., $\bar{n}=200,400,600,800,1000$), and the arrow direction shows an increase with squeezing.
		The white dashed line shows the minimum of the rescaled QFI versus displacement and squeezing.
		Other parameters are the same as those in Fig.~\ref{Fig2-HL-CS}.
	}\label{Fig3-DensityPlotQFI-DSVS}
\end{figure}

\emph{Displaced squeezed vacuum state}.---In current realistic experimental capability, it is, however, hard to prepare a perfect squeezed photon state.
It often involves the simultaneous coexistence of both squeezing and displacement.
To elucidate the distinct roles played by the optical displacement and the optical squeezing, we consider the initial state is a product state $\left|\psi\right\rangle =\left|\alpha,\xi\right\rangle \otimes\left|\mu\right\rangle $, where the optical part is a displaced squeezed vacuum state (DSVS), i.e. $\left|\alpha,\xi\right\rangle =D\left(\alpha\right)S\left(\xi\right)\left|0\right\rangle $ with $n_{\text{th}}=0$. 
In this scenario, the average photon number is given by 
$\bar{n}  =\left|\alpha\right|^{2}+\sinh^{2}r$ and we define $\beta $ as the ratio of quantum squeezing and the amplitude of the coherent state, i.e., $\beta \equiv \sinh^{2}r/ \left|\alpha\right|^{2}$.


Now, we explore to what extent quantum squeezing can ensure the realization of  the double-HS precision.
From the Eq.~\eqref{Vara+a}, we obtain the photon-number fluctuation
\begin{align}
	\text{Var}\!\left(\!a^{\dagger}a\!\right) 
	& \!=\! \frac{1}{2}\sinh^{2}2r+\left|\alpha\right|^2\left[\cosh 2r-\sinh 2r\cos\left(2\zeta-\vartheta\right)\right]\nonumber\\
	&\!=\!\frac{1}{2}\sinh^{2}\!2r\!+\!\frac{\bar{n}}{1+\beta}\left(\cosh 2r-\tau\sinh 2r\right), 
\end{align}
where $\tau\equiv \cos\left(2\zeta-\vartheta\right)$ reflects the phase configuration of the initial optical state.
Given the fact that $2\sinh^{2}r\leq\sinh2r<2\sinh^{2}r+1$, it follows that
\begin{align}\label{2}
	\text{Var}\left(a^{\dagger}a\right) 
	& >  \text{Var}_{+}\nonumber \\
	& \equiv \frac{\bar{n}}{1+\beta}\left[ \frac{2\beta\bar{n}}{1+\beta}\left(\beta+1-\tau\right)+1-\tau\right]
\end{align}
for $\tau\geq0$, and 
\begin{align}\label{3}
	\text{Var}\left(a^{\dagger}a\right) 
	& \geq  \text{Var}_{-} \nonumber\\
	& \equiv \frac{\bar{n}}{1+\beta}\left[ \frac{2\beta\bar{n}}{1+\beta}\left(\beta+1-\tau\right)+1\right] 
\end{align}
for $\tau<0$.
When $\beta\gg1/\left(2\bar{n}\right)$, Eqs.~\eqref{2} and~\eqref{3} further suggest the scaling of quantum fluctuation 
\begin{align}
	\text{Var}\left(a^{\dagger}a\right) & \gtrsim \frac{\bar{n}}{1+\beta}\left(\frac{2\beta^{2}\bar{n}}{1+\beta}+1\right)
	\gtrsim \frac{2\beta^{2}}{\left(1+\beta\right)^{2}}\bar{n}^{2}.
\end{align}	
Based on the above analysis, we have $\text{Var}\left(a^{\dagger}a\right)\approx2\bar{n}^{2}$, $\left\langle \mu\right|J_{z}^{2}\left|\mu\right\rangle =N^{2}/4$, and $\text{Var}\left(J_{z}\right)=0$ for a spin-coherent state with a large enough $\beta$ and the angle $\theta=0$.
Consequently, the QFI~\eqref{F_DSTS} gives 
\begin{equation}
	\mathcal{F}_h^{\text{DSVS}} \! =  \!\frac{4g^{4}}{\Delta^{4}}t^{2}N^{2}\text{Var}\left(a^{\dagger}a\right)
	\!\gtrsim\!\frac{8g^{4}}{\Delta^{4}}\frac{\beta^{2}}{\left(\!1\!+\!\beta\!\right)^{2}}t^{2}N^{2}\bar{n}^{2},
\end{equation}
showing a novel existence of the double-HS. 
It is noticed that we also require the average photon and atom numbers satisfying  $\Delta\gg g\sqrt{N}\gg g\sqrt{N^{2}/\bar{n}}$ (corresponding to the requirement in the time-averaged method) and the squeezing satisfying  $\beta\gg1/\left(2\bar{n}\right)$. 
We observe that the growth of squeezing can enhance measurement precision for a fixed number of atoms.

We now further ascertain the conditions under which the QFI increases with the quantum squeezing.
In Fig.~\ref{Fig3-DensityPlotQFI-DSVS}, the density plot depicts the rescaled QFI $\ln\left(\mathcal{F}_h^{\text{DSVS}}/t^{2}\right)$, as a function of displacement $\left|\alpha\right|^{2}$ and squeezing $\sinh^{2}r$, with a fixed $\tau=1$.
For a given average photon number $\bar{n}$ (indicated by the grey dot-dashed lines), the QFI first decreases and subsequently increases as the squeezing component rises.
The white dashed line delineates the minimum of the QFI. 
We observe that, at least for the $\tau=1$ case, squeezing is not always beneficial to the QFI for a small value of squeezing.

\begin{figure}[tbp]
	\centering
	\begin{minipage}{1\linewidth}
		\begin{overpic}[width=\linewidth]{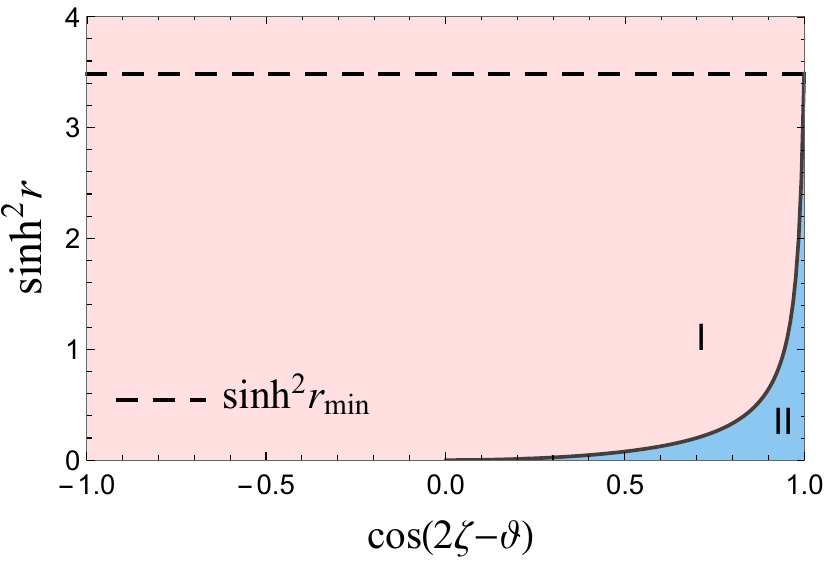}\label{fig4-a}
			\put(0,65){(a)}
		\end{overpic}
	\end{minipage}\\
	\begin{minipage}{1\linewidth}
		\begin{overpic}[width=\linewidth]{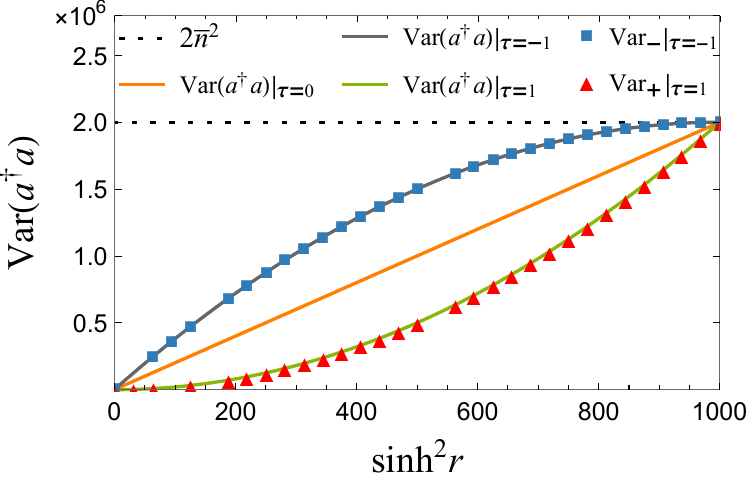}\label{fig4-b}
			\put(0,62){(b)}
		\end{overpic}
	\end{minipage}
	\protect\protect\caption{Monotonicity of the QFI $\mathcal{F}_{h}^{\text{DSVS}}$ and the photon-number fluctuation $\text{Var}\left(a^{\dagger}a\right)$ for DSVSs.
		(a) Monotonic regions of the QFI $\mathcal{F}_{h}^{\text{DSVS}}$ versus the squeezing $\sinh^{2}r$ and the angle condition $\tau=\cos\left(2\zeta-\vartheta\right)$ with a given average photon number $\bar{n}=1000$.
		The black line indicates the derivative zeros $\partial_{\sinh^{2}r}\mathcal{F}_{h}^{\text{DSVS}}=0$, dividing the parameter space into two regions: monotonically increasing region-I (pink) and monotonically decreasing region-II (cyan).		
		The black dashed line $\sinh^{2}r_{\text{min}}$ denotes the squeezing value at the boundary when $\tau=1$.
		(b) The photon-number fluctuation $\text{Var}\left(a^{\dagger}a\right)$ as a function of the squeezing $\sinh^{2}r$ with a fixed average photon number $\bar{n}=1000$ and specific angle conditions $\tau=0, \pm 1$, respectively.
		The fluctuations for $\tau=-1$ (gray line), $\tau=0$ (orange line), $\tau=1$ (green line), their approximations $\text{Var}_{-}|_{\tau=-1}$ (blue squares) and $\text{Var}_{+}|_{\tau=1}$ (red triangles), and the maximum fluctuation $2\bar{n}^{2}$ (black dots), confirm the scalings Eqs.~\eqref{2} and~\eqref{3}.
		%
		%
		Other parameters are the same as those in Fig.~\ref{Fig2-HL-CS}.
	}
	\label{Fig4}
\end{figure}

In Fig.~\ref{Fig4}a, the QFI $\mathcal{F}_{h}^{\text{DSVS}}$ depicts the role of squeezing parameter in terms of $\sinh^{2}r$ and the phase parameter $\tau=\cos\left(2\zeta-\vartheta\right)$ for a fixed average photon number $\bar{n}=1000$. 
The parameter space can be divided into two distinct regions: the QFI exhibits a monotonic increase with growing squeezing in region-I, while the QFI experiences a monotonic decrease as squeezing increases in region-II.
%
Therefore, for $\tau<0$, regardless of the squeezing value, an increase in the squeezing always leads to a growth of the QFI.
Meanwhile, for $\tau<0$,   the phase parameter provides an additional positive effect to the QFI $\mathcal{F}_{h}^{\text{DSVS}}=4g^{4}t^{2}N^{2}\text{Var}\left(a^{\dagger}a\right)/\Delta^{4}$, see Fig.~\ref{Fig4}b.
We thus can optimize the efficiency of the quantum resource (squeezing) to maximize the measurement precision by controlling the phase parameter ($\tau=-1$) in the presence of an imperfect squeezing light.
Therefore, for a DSVS, the double-HS precision is attainable if the ratio $\beta$ between the quantum squeezing and the amplitude of the coherent state is large enough $\beta\gg 1/(2\bar n)$. 
%

\subsection{Optical coherent state with a spin-squeezed state}
We have demonstrated the realization of a double-HS precision by building on optical squeezing. 
Now we further explore the role of spin squeezing in the metrological sensibility of measuring the magnetic field. 
To this end, we consider an initial state $\left|\psi\right\rangle =\left|\alpha\right\rangle \otimes\left|\phi\right\rangle $, comprising a classical coherent state $\left|\alpha\right\rangle$ for the optical component and a nonclassical spin-squeezed state~\cite{Kitagawa1993,MA201189,PhysRevA.81.022106,PhysRevA.102.052423} for the atomic component
\begin{align}
	\left|\phi\right\rangle  & = \text{e}^{-\text{i}\frac{\chi}{2}J_{x}^{2}}\left|1\right\rangle.
\end{align}
Here, $\left|1\right\rangle =\left|j,-j\right\rangle$ is the collective ground state with the eigenvalue $-N/2$, $\chi=2\kappa t$ is the one-axis twisting angle and $\kappa$ is the coupling constant.
Then we have the fluctuation
\begin{align}
	\text{Var}\left(J_z\right)&=\frac{1}{8}N\left[\left(N\!-\!1\right)\cos^{N-2}\chi-2N\cos^{2N-2}\left(\frac{\chi}{2}\right)\right. \nonumber \\
	&\quad\left.+N+1\right]. 
\end{align}
For the one-axis twisting angle $\chi=\pi+2k\pi$ ($k$ is integer), we have 
\begin{equation}
	\text{Var}\left(J_{z}\right)
	=\begin{cases}
		\frac{N^{2}}{4}, & \ \text{even}\ N;\\
		\frac{N}{4}, & \ \text{odd}\ N. 
	\end{cases}
\end{equation}
Moreover, for $\chi=\pi/2+k\pi$ and the spin number $N\gg 1$, we find
$\text{Var}\left(J_z\right) \approx  N^{2}/8$ is insusceptible to the parity of spin $N$.
Here, we may chose the case $\chi=\pi+2k\pi$ with an even $N$ and $\text{Var}\left(a^{\dagger}a\right)=\left|\alpha\right|^{2}=\bar{n}\gg\Delta^{2}/\left(2g^{2}\right)$ so that 
the maximum QFI is given by 
\begin{align}\label{main-result-2}
	\mathcal{F}_h 
	&= 4t^{2}\left[\left(1-\frac{2g^{2}}{\Delta^{2}}\bar{n}\right)^{2}\text{Var}\left(J_{z}\right)+\frac{4g^{4}}{\Delta^{4}}\text{Var}\left(a^{\dagger}a\right)\left\langle J_{z}^{2}\right\rangle \right]\n 
	&\gtrsim  \frac{4g^{4}}{\Delta^{4}}t^{2}N^{2}\bar{n}^{2}. 
\end{align}
By comparison, we observe that both Eq.~\eqref{O-SVS} and  Eq.~\eqref{main-result-2} give a double-HS precision. 
In the latter,  the utilization of atom-number fluctuation results in  $\mathcal{F}_h\propto \bar n^2{\rm Var}(J_z)$.
%
Similarly, we could also obtain the maximum QFI $\mathcal{F}_{h}\approx12g^{4}t^{2}N^{2}\bar{n}^{2}/\Delta^{2}$  in frequency estimation by inducing both the light and spin squeezing with $\chi=\pi+2k\pi$, $\left|\alpha\right|=n_{\text{th}}=0$ and large even $N$. 
%

\section{Conclusion and Discussion}\label{SEC5}
In a prototypical and theoretically significant cavity QED, TC model, we have explored the feasibility of achieving double-HS precision. 
The derived effective Hamiltonian~\eqref{H_eff} unveils the emergence of the coherent averaging mechanism even in estimating a non-global parameter, which leads to the HS precision for the photon number when classical states serve as the probing entities.
The underlying cause of the HL scaling resides in the fact that the evolved atom state (i.e., quantum bus) captures the accumulated phase arising from the interaction between the cavity field and the quantum bus.
%
Hence, the HL scaling of our scheme could be achieved by measuring the observables of the quantum bus or the entire system even without nonclassical inputs, as long as they do not commute with the interaction Hamiltonian $2g^{2}J_{z}a^{\dagger}a/\Delta$~\cite{braun_heisenberg-limited_2011,https://doi.org/10.1002/andp.201500169}.

Considering nonclassical inputs, we have derived an analytical expression for the QFI, showing that the double-HS precision is attainable through the introduction of either the photon-number fluctuation via optical squeezing or the atom-number fluctuation via spin squeezing.
In addition, we have also investigated different metrological scenarios, where optical squeezing and displacement coexist. 
In such instances, we have observed that an increase in the squeezing can decrease measurement precision when the photon number remains fixed and squeezing is small.
Judiciously adjusting the phase parameter could offer a viable solution to mitigate the imperfect squeezing.
%

%
Our work provides deeper insights into the realization of the double-HS precision in the models with star-like long-range interactions. 
Our findings hold immediate implications for determining the minimum quantum resource prerequisites for high-precision metrology in more intricate metrological networks featuring multiple quantum buses.
Recent remarkable experimental advancements in attaining strong coupling~\cite{WOS:000243867300038,WOS:000322086100035,PhysRevA.71.013817,PhysRevA.67.033806,WOS:000257665300034,PhysRevLett.128.123602,PhysRevLett.130.173601}, as well as the development of the TC model~\cite{PhysRevLett.103.083601}, also instill optimism regarding the prospective realization of our theoretical framework.

\begin{acknowledgments}
This work is supported by the National Natural Science Foundation of China (Grants No. 12247158, and No. 12405026), the National Natural Science Foundation of China Key (Grants No. 92365202, and No. 12134015), the Hangzhou Joint Fund of the Zhejiang Provincial Natural Science Foundation of China under Grant No. LHZSD24A050001, the Funds of the Natural science Foundation of Hangzhou under Grant No. 2024SZRYBA050001, the ``Wuhan Talent'' (Outstanding Young Talents), and the Postdoctoral Innovative Research Post in Hubei Province.
\end{acknowledgments}

\section*{Data Availability Statement}

The data that support the findings of this study are available from the corresponding author upon reasonable request.

\section*{References}
\nocite{*}

%

\end{document}